\documentclass[12pt,a4paper,epsf]{article}
\usepackage{graphics}
\usepackage{amssymb,amsmath}
\usepackage[dvips]{lscape,graphicx}
\usepackage{cite}
\usepackage{longtable}

\newcommand{\ct}{\cite}
\newcommand{\lb}{\label}

\newcommand{\bc}{\begin{center}}
\newcommand{\ec}{\end{center}}
\newcommand{\bd}{\begin{displaymath}}
\newcommand{\ed}{\end{displaymath}}
\newcommand{\be}{\begin{equation}}
\newcommand{\ee}{\end{equation}}
\newcommand{\ba}{\begin{array}}
\newcommand{\ea}{\end{array}}
\newcommand{\bea}{\begin{eqnarray}}
\newcommand{\eea}{\end{eqnarray}}
\newcommand{\bt}{\begin{tabular}}
\newcommand{\et}{\end{tabular}}
\newcommand{\un}{\underline}

\newcommand{\bp}{\begin{picture}}
\newcommand{\ep}{\end{picture}}
\newcommand{\bfi}{\begin{figure}}
\newcommand{\efi}{\end{figure}}

\def\fun#1#2{\lower3.6pt\vbox{\baselineskip0pt\lineskip.9pt
\ialign{$\mathsurround=0pt#1\hfil##\hfil$\crcr#2\crcr\sim\crcr}}}

\begin{document}

 \vspace{1cm}

\title{\large \bf {Superstring-Inspired
$\large \bf E_6$ Unification,\\ Shadow Theta-Particles and
Cosmology }}
\author{\bf C.R.~Das \\
\large Centre for Theoretical Particle Physics, Technical
University of Lisbon, \\ Avenida Rovisco Pais, 1 1049-001 Lisbon,
Portugal\\[5mm]
\un{\bf L.V.~Laperashvili} \\
The Institute of Theoretical and Experimental Physics,\\
Bolshaya Cheremushkinskaya, 25, 117218 Moscow, Russia\\[5mm]
and
\bf A.~Tureanu \\
Department of Physics, University of Helsinki\\ and Helsinki
Institute of Physics, P.O.Box 64, FIN-00014 Helsinki, Finland}

\date{}

\maketitle

E-mail:{\,\,crdas@cftp.ist.utl.pt,\,\,laper@itep.ru,\,\,
anca.tureanu@helsinki.fi}

 \vspace{1cm}

\bc {\bf A talk given at the International Bogolyubov Conference}\\
{\bf "Problems of Theoretical and Mathematical Physics",}\\ {\bf
Dubna, Joint Institute for Nuclear Research (JINR), August 23-27,
2009.}\ec

\vspace{0.5cm}

\bc Published in Physics of Particles and Nuclei, 2010, Vol. 41,
No. 6, pp. 965–-968. \ec

\vspace{1cm}

\begin{center}{\bf Abstract}\end{center}
\begin{quote}

We construct a new cosmological model considering the
superstring-inspired $E_6$ unification in the 4-dimensional space
at the early stage of the Universe. We develop a concept of
parallel existence in Nature of the ordinary and  shadow worlds
with different cosmological evolutions.

\end{quote}

\section{Introduction}

We have developed a concept of parallel existence of the ordinary
(O) and  shadow (Sh) worlds assuming the superstring-inspired
$E_6$ unification in the 4-dimensional space \ct{1}. The
'heterotic' superstring theory $E_8\times E'_8$ was suggested as a
more realistic model for unification of all fundamental gauge
interactions with gravity \ct{2}. This ten-dimensional theory can
undergo spontaneous compactification. The integration over six
compactified dimensions of the $E_8$ superstring theory leads to
the effective theory with the $E_6$ unification in the
four-dimensional space. Superstring theory has led to the
speculation that there may exist another form of matter - ``shadow
matter'' - in the Universe, which only interacts with ordinary
matter via gravity, or gravitational-strength interactions \ct{3}.
The shadow world, in contrast to the mirror world \ct{4}, can be
described by another group of symmetry (or by a chain of groups of
symmetry), which is different from the ordinary world symmetry
group.

Our model is based  on the following assumptions:\\
$\bullet$ Grand Unified Theory (GUT) is inspired by Superstring
theory \ct{2}, which predicts the existence of the $E_6$
unification in the 4-dimensional space at the energy scale
$\sim 10^{18}$ GeV.\\
$\bullet$ The Shadow world is responsible for the dark energy (DE)
and dark matter (DM).\\
$\bullet$ We assume that $E_6$ unification had a place in the O-
and M- worlds at the early stage of our Universe. This means that
at the very high energy scale $\sim 10^{18}\,\,{\rm{GeV}}$ there
exist mirror world (MW) and the group of symmetry $E_6\times E'_6$
\ct{1}.  \\
$\bullet$ We have adopted for the O-world the breaking
         $E_6\to SO(10)\times U(1),$
while for the Sh-world we have considered the breaking $E'_6\to
SU(6)'\times SU(2)' $ with aim to have in the Sh-world an extra
$SU(2)_\theta'$ group at low energies. Here ordinary (shadow)
world is described by non-primed (primed) symbols.

The study \ct{1} is a development of the ideas considered
previously in Refs.\ct{5}. However, in present investigation we
give a new interpretation of the possible accelerating expansion
of the Universe using a cosmological quintessence model with
superstring-inspired $E_6$ unification .

\section{$\bf E_6$ unification breakdown in the ordinary and shadow
worlds}

Developing the ideas of Refs.~\ct{1}, we have considered the
existence of theta-particles in the shadow world. These
"theta-particles" were introduced by Okun in Refs. \ct{6}, who
suggested that there exists in Nature the symmetry group
\be SU(3)_C\times SU(2)_L\times SU(2)_{\theta}\times U(1)_Y   \,,
\lb{27} \ee
which, in contrast to the Standard Model (SM) group, has an
additional non-Abelian $SU(2)_{\theta}$ group whose gauge fields
are neutral massless vector particles -- 'thetons'. These
'thetons' have a macroscopic confinement radius
$1/\Lambda_{\theta}$. In Refs. \ct{1} we have assumed that such a
group of symmetry exists in the shadow world at low energies and
having $\Lambda'_{\theta}\sim 10^{-3}$ eV  provides a tiny
cosmological constant.

According to the assumptions of Refs.~\ct{1}, in the ordinary
world, from the SM-scale up to the $E_6$ unification scale, there
exists the following chain of symmetry groups:
$$SU(3)_C\times SU(2)_L\times U(1)_Y
\to  [SU(3)_C\times SU(2)_L\times U(1)_Y]_{{SUSY}}
$$ $$\to
 SU(3)_C\times SU(2)_L \times SU(2)_R\times U(1)_X\times
 U(1)_Z\to SU(4)_C\times SU(2)_L \times SU(2)_R\times U(1)_Z $$ \be \to
 SO(10)\times U(1)_Z \to E_6. \lb{34o} \ee
But the following chain of symmetry groups exists in the shadow
world:
$$
SU(3)'_C\times SU(2)'_L\times SU(2)'_{\theta}\times U(1)'_Y \to
[SU(3)'_C\times SU(2)'_L\times SU(2)'_{\theta}\times
U(1)'_Y]_{{SUSY}}
$$ $$\to SU(3)'_C\times
SU(2)'_L\times SU(2)'_{\theta}\times U(1)'_X \times U(1)'_Z\to
SU(4)'_C\times SU(2)'_L\times SU(2)'_{\theta}\times U(1)'_Z $$ \be
\to SU(6)'\times SU(2)'_{\theta}\to E'_6.
                                          \lb{34sh} \ee
\section{New shadow gauge group $\bf SU(2)'_{\theta}$}

The reason
for our choice of the $SU(2)'_{\theta}$ group was to obtain a new
scale in the shadow world at extremely low energies \ct{1},\ct{5}.
By comparison with the content of the 27-plet of $E_6$, having 16
fermions (see \ct{1}), we have considered theta-quarks as
$\theta-$doublets and shadow leptons as $\theta-$singlets. The
scalars $\phi'_{\theta}$ also can be considered as doublets of
$SU(2)'_{\theta}$.

In Refs.~\ct{1} we have considered the running coupling constant
${\alpha'}_{2\theta}^{-1}(\mu)$ for high energies $\mu
> M'_t$ (where $M_t'$ is the top-quark mass in the Sh-world),
assuming the existence of three generations of theta-quarks and
two doublets of scalar fields $\phi'_{\theta}$. Of course, near
the scale  $\Lambda'_{\theta}$ only theta-quarks of the first
generation contribute, and it is easy to obtain the value
$\Lambda'_{\theta}\sim 10^{-3}$ eV. Theta-quarks of the first
generation are stable, due to the conservation of the theta-charge
\ct{6}. We also have considered a complex scalar field
$\varphi_{\theta} = (1,1,1,0)$, which is a singlet under the
symmetry group $G' = SU(3)'_C\times SU(2)'_L\times
SU(2)'_{\theta}\times U(1)'_Y.$

\section{Cosmological Constant, DE and DM}

>From the point of view of particle physics the cosmological
constant naturally arises as an energy density of the vacuum. For
the present epoch, the Hubble parameter $H$ is given by the
following value \ct{7}: $ H = 1.5 \times 10^{-42}\,\,{\rm{GeV}}$,
and the critical density of the Universe is
\be \rho_{c} = 3H^2/8\pi G = {(2.5\times 10^{-12}\,\,
{\rm{GeV}})}^4. \lb{31} \ee
For the ratios of densities $\Omega_X = \rho_x/\rho_c$,
cosmological measurements give the following density ratios of the
total Universe \ct{7}: $\Omega_0 = \Omega_r + \Omega_M +
\Omega_\Lambda = 1$. Here $\Omega_r$ is a relativistic (radiation)
density ratio, and $\Omega_{\Lambda} = \Omega_{DE}$. The
measurements give: $\Omega_{DE}\sim 75\%$ - for the mysterious
dark energy, $\Omega_M \approx \Omega_B + \Omega_{DM} \sim 25\%$,
$\Omega_B \approx 4\%$ - for (visible) baryons, $\Omega_{DM}
\approx 21\%$ - for dark matter. Here we propose that a plausible
candidate for DM is a shadow world with its shadow quarks,
leptons, bosons and super-partners, and the shadow baryons are
dominant:
 $\Omega_{DM} \approx \Omega_{B'}.$
Then we see that
$\Omega_{B'} \approx 5\Omega_{B},$
what means that the shadow baryon density is larger than the
ordinary baryon density.

We can calculate the dark energy density using the results given
by \ct{7}:
\be \rho_{DE} \approx 0.75\ \rho_c \approx (2.3 \times 10^{-3}
\,\, {\rm eV})^4. \lb{32} \ee
The $\Lambda CDM$-cosmological model  predicts that the
cosmological constant $\Lambda$ is equal to $ \Lambda =
\rho_{vac}= \rho_{DE}.$ Given by Eq.~(\ref{32}), the cosmological
constant $\Lambda$ is extremely small:
\be \Lambda \approx (2.3 \times 10^{-3} \,\, {\rm eV})^4. \lb{33}
\ee
The evolution of the Universe is described by an equation of
state:
\be   w = \frac{p}{\rho}, \lb{20} \ee
where $w$ in general is assumed to be constant, but may be
time-dependent. If the cosmological constant describes the dark
energy (DE), then  $\rho_{DE}= - p_{DE},$ and $w = w_{DE}= -1 $.
Recent cosmological observations \ct{7} give the following value
of $w$: $ w = -1.05\pm 0.13\,\,{\rm{(statistical)}}\pm 0.09\,\,
{\rm{(systematic)}}$.

\section{Quintessence model of DE, DM and matter}

The main idea of our present investigation is to show the absence
of the SM and SM' contributions to the cosmological constant
$\Lambda$. We relate the value (\ref{33}) only with the
$SU(2)'_{\theta}$ gauge group's contribution.

We assume that there exists an axial $U(1)_A$ global symmetry in
our theory with currents having $SU(2)'_\theta$, $SU(3)_c$ and
$SU(3)'_c$ anomalies, which are spontaneously broken at the scale
$f$ by a singlet complex scalar field $\varphi$, with a VEV
$\langle \varphi \rangle = f$. As a result, we have three
Nambu-Goldstone (NG) bosons:
\be \varphi^{(i)} = (f + \sigma^{(i)}) \exp(i \alpha^{(i)}/f),
\quad i=0,1,2, \lb{52} \ee
where index $i=0,1,2$ corresponds to $SU(2)'_\theta$, $SU(3)'_c$
and $SU(3)_c$ gauge groups, respectively. The boson $\alpha^{(i)}$
(imaginary part of the singlet scalar field $\varphi^{(i)}$) is an
axion and could be identified with a massless Nambu-Goldstone (NG)
boson if the $U(1)_A$ symmetry is not spontaneously broken.
However, the spontaneous breaking of the global $U(1)_A$ by
corresponding instantons inverts $\alpha^{(i)}$ into the pseudo
Nambu-Goldstone (PNG) bosons.

The singlet complex scalar field $\varphi^{(i)}$ reproduces a
Peccei-Quinn (PQ) model \ct{8}.  Near the vacuum, a PNG mode
$\alpha^{(i)}$ emerges the following PQ axion potential:
\be V_{PQ}(\alpha^{(i)}) \approx {(\Lambda^{(i)})}^4
         \left(1 - \cos(\alpha^{(i)}/f)\right).  \lb{53} \ee
This axion potential exhibits minima at
\be \cos(\alpha^{(i)}/f) = 1,\quad \rm{i.e.}\quad
{(\alpha^{(i)})}_{min}= 2\pi n f, \quad n = 0,1,... \lb{55} \ee
For small fields $\alpha^{(i)}$ we expand the effective potential
(\ref{53}) near the minimum:
\be    V_{PQ} \approx \frac{({\Lambda^{(i)})}^4}{2f^2}
(\alpha^{(i)})^2 + ... = \frac 12 m_i^2 {(\alpha^{(i)})}^2 + ...,
\lb{56} \ee
and hence the PNG axion mass squared is given by:
\be m_i^2\sim {(\Lambda^{(i)})}^4/f^2.  \lb{57} \ee
Let us assume that at the cosmological epoch when $U(1)_A$ was
spontaneously broken, the value of the axion fields $\alpha^{(i)}$
was deviated from zero and initial values were:
$\alpha_{in}^{(i)}\sim f$. The scale $f\sim 10^{18}$ GeV (near the
$E_6$ unification scale) makes it natural that the $U(1)_A$
symmetry was broken before inflation.

Now quintessence is described by the PNG scalar field $\alpha
\equiv \alpha^{(i)}$ minimally coupled to gravity and leading to
the late time inflation. The action for quintessence is given by
\be  S = \int d^4x \sqrt{-g}[\frac{1}{2} {(\nabla \alpha)}^2 -
V(\alpha)], \lb{58} \ee
where $V(\alpha)$ is the potential of the field. In the flat
Friedmann spacetime the action (\ref{58}) leads to the following
field and Einstein equations:
\be  \ddot{\alpha} + 3H\dot{\alpha} + \frac{dV}{d\alpha} = 0.
\lb{59} \ee
\be H^2 = \frac{8\pi G}{3}[\frac 12 (\dot{\alpha})^2 + V(\alpha)],
 \lb{65}     \ee
\be \frac{\ddot{a}}{a} = - \frac{8\pi G}{3}[\frac 12
(\dot{\alpha})^2 - V(\alpha)].
  \lb{66}  \ee
The equation of state is given by
\be w^{(i)} = \frac{(\dot{\alpha})^2 -
2V(\alpha^{(i)})}{(\dot{\alpha})^2 + 2V(\alpha^{(i)})}. \lb{67}
\ee

\section{Quintessence model for $SU(2)'_\theta$ gauge group}

This is a case for $i=0$. Solving Eq.~(\ref{59}) for
$\alpha^{(0)}$  we can introduce an axion ${\alpha}_\theta$ by the
notation $\alpha^{(0)}={\alpha}_\theta$.
The equation of motion (EOM) of the classical field
${\alpha}_\theta$ is:
\be \frac{d^2 {\alpha}_\theta}{dt^2} + 3 H \frac{d
{\alpha}_\theta}{dt} + V'({\alpha}_\theta) = 0, \lb{69} \ee
where according to Eq.~(\ref{53}),
$$ V({\alpha}_\theta) =
{(\Lambda'_\theta)}^4
         \left(1 - \cos({\alpha_\theta}/f)\right), $$
\be V'({\alpha}_\theta)
=\frac{{(\Lambda'_\theta)}^4}{f}\sin(\alpha_\theta/f). \lb{71} \ee
If now $\sin(\alpha_\theta/f)=0$, but $\cos(\alpha_\theta/f)=-1$,
then $\dot{\alpha}_\theta =0$, and
\be V(\alpha_\theta) = 2{(\Lambda'_\theta)}^4.  \lb{72} \ee
In this case ${\alpha}_\theta={\rm{const}}$,
${\rho}_\theta={\rm{const}}$, $w=w_\theta=-1$, what means that
axion $\alpha^{(0)}=\alpha_\theta$ gives the contribution to DE
with DE density equal to the value (\ref{72}):
\be \rho_{DE} = 2{(\Lambda'_\theta)}^4.   \lb{73} \ee
Using the result (\ref{32}), we obtain the following estimate of
the $SU(2)'_\theta$ group's gauge scale:
\be \Lambda'_\theta \backsimeq   2.0\times 10^{-3}\,\,\rm{eV}  .
\lb{74} \ee
If $\Lambda'_\theta \sim 10^{-3}$ eV and $f\sim 10^{18}$ GeV, then
from Eq.~(\ref{57}) we obtain the estimate of the theta-axion
mass: $m_0\sim {\Lambda'_{\theta}}^2/f\sim 10^{-42}
\,\,{\rm{GeV}}$.

\section{Quintessence model for $SU(3)_c$ and $SU(3)'_c$ gauge
groups}

In the case $i=1,2$ we have $\Lambda^{(2)}=\Lambda_{QCD}\sim 0.3$
GeV and $\Lambda^{(1)}=\Lambda'_{QCD}\approx 1.5\Lambda_{QCD}$
(see \ct{1}), which are much larger than $\Lambda'_\theta$. We can
consider $\dot{\alpha}^{(1,2)}\neq 0$ if
$V'=dV/d\alpha^{(1,2)}\neq 0$ in Eq.~(\ref{59})  .

Assuming that at present epoch we have ${(\dot{\alpha}^{(1,2)})}^2
=  2V(\alpha^{(1,2)})$,
we obtain from  Eq.~(\ref{67}): $w^{(1,2)}=w_m = 0$,
what describes the matter contributions to $\rho_{(QCD)}$ and
$\rho'_{(QCD)}$ with the result:
\be \frac{\rho_{DM}}{\rho_B}\simeq
\left(\frac{\Lambda'_{QCD}}{\Lambda_{QCD}}\right)^4 = {(1.5)}^4
\simeq 5. \lb{87} \ee

\end{document}